\documentclass[aps,twocolumn,prl,showpacs]{revtex4}
\usepackage{epsfig}
\usepackage{latexsym}
\begin{document}      
\title{A lattice mesoscopic model of dynamically
heterogeneous fluids}
\author{A. Lamura}
\affiliation{
Istituto Applicazioni Calcolo, CNR, Sezione di Bari,
Via Amendola 122/D, 70126 Bari, Italy}
\author{S. Succi}
\affiliation{
Istituto Applicazioni Calcolo, CNR, 
V.le del Policlinico 137, 00161 Roma, Italy
}
\date{\today}
\begin{abstract}
We introduce a mesoscopic three-dimensional Lattice Boltzmann Model which 
attempts to mimick the physical features 
associated with cage effects in dynamically heterogeneous fluids.
To this purpose, we extend the standard Lattice Boltzmann 
dynamics with self-consistent constraints based on the 
non-local density of the surrounding fluid.
The resulting dynamics exhibits typical features of dynamic
heterogeneous fluids, such as non-Gaussian
density distributions and long-time relaxation.  
Due to its intrinsically parallel dynamics, and absence of statistical
noise, the method is expected to compute significantly 
faster than molecular dynamics,
Monte Carlo and lattice glass models.
\end{abstract}
\maketitle

Very slow relaxation and long equilibration times are a
typical feature of complex fluids, such as 
supercooled liquids, polymers and others \cite{review,garr}.
The numerical study of these complex systems is usually
undertaken by many-body simulation methods, such as
Molecular Dynamics \cite{md}, Monte Carlo \cite{mc},
and various types of lattice 'glasses' \cite{KA,lg,coniglio,rit,ton,bertin}. 
Since many-body simulations are computationally
demanding, it is worth exploring whether the salient features
of dynamic heterogeneities can be reasonably described
by effective one-body techniques in the spirit of
density functional theory \cite{dft}.
A powerful one-body technique is the Lattice Boltzmann (LB) method.
The LB dynamics consists of three basic steps:
Free-streaming, collisional relaxation \cite{HSB}
and (effective) intermolecular interactions \cite{lbe-1}.
Judicious choice of these intermolecular interactions
permits to describe the dynamics of a variety of complex flows \cite{lbe-2}.
However, applicability of LB to glassy-like fluids 
remains an open problem \cite{noi}.
A crucial aspect of the physics of complex fluids are dynamic heterogeneities  
and geometrical frustration, i.e. the presence of sterical 
constraints which reduce the phase-space available to the fluid system.
Real systems like
colloids, or granular materials, exhibit glassy dynamics associated
to jamming: Density, rather than temperature, is the dominant effect
in slowing down strongly the dynamics \cite{ediger}.
In an attempt to model these effects, we develop a lattice Boltzmann 
equation (LBE) in which free-particle motion is confined to
a subset of links which fulfill self-consistent dynamic constraints on the
surrounding fluid density. 
The presence of these kinetic constraints on free
particle motion leads to drastic departures from 
simple fluid behavior, such as long-time relaxation
and non-Gaussian density fluctuations.
We begin by considering a standard LBE
with a single relaxation time \cite{bhat54,qian}:
\begin{equation} 
f_i({\bf r},t) -
f_i({\bf r}_i,t-\Delta t) =
-\omega \Delta t \left [ f_i-f_i^e \right ] ({\bf r}_i,t-\Delta t) 
\label{lbe}
\end{equation} 
where $\Delta x$ and $\Delta t$ are space- and time-steps, respectively,
$f_i({\bf r},t) \equiv f({\bf r},{\bf v}={\bf c}_i,t)$ is a
discrete distribution function of particles at site ${\bf r}$ and at time $t$
moving along the direction $i$ of a lattice, 
with discrete speed ${\bf c}_i$, and 
${\bf r}_i \equiv {\bf r}-{\bf c}_i \Delta t$.
The previous equation can be seen as the combination of 
collision and streaming steps. 
In the collision, the distribution functions
$f_i$ relax to a local equilibrium $f_i^e$
in a time lapse of the order of $\omega^{-1}$, such that the distribution
function after a collision $f_i^c$ is:
\begin{equation} 
f_i^c({\bf r}_i,t-\Delta t) =
f_i({\bf r}_i,t-\Delta t) 
-\omega \Delta t \left [ f_i-f_i^e \right ] ({\bf r}_i,t-\Delta t) 
\label{lbe1}
\end{equation} 
The distribution function $f_i^c$ then streams freely 
to neighbor sites, so that the updated values are
just the shifted post-collisional distributions:
\begin{equation} 
f_i({\bf r},t) = f_i^c({\bf r}_i,t-\Delta t) 
\label{lbe2}
\end{equation} 
The large-scale behavior of system depends crucially
on the form of the local equilibria.
In the present work, the equilibrium distribution 
functions $f_{i}^{e}$ are expressed as:
\begin{equation}
f_i^e({\bf r},t) = w_i \rho({\bf r},t) 
\label{fequil}
\end{equation}
where $w_i$ is a set of lattice-dependent weights normalized to unity.
The local density $\rho({\bf r},t)$ in eq.~(\ref{fequil}) 
is obtained by a direct summation upon all 
discrete distributions:
\begin{equation}
\rho({\bf r},t)  =  \sum_{i} f_{i}({\bf r},t).
\label{dens}
\end{equation}
Since the local equilibria do not depend on the local
fluid speed, the only conserved quantity is the fluid density, which
means that in the continuum limit, the system obeys a
simple diffusion equation:
$\partial_t \rho({\bf r},t) = D \nabla^2 \rho({\bf r},t)$
with diffusion constant 
$D= \frac{(\Delta x)^2}{3 \Delta t}
\left ( \frac{1}{\omega \Delta t} - \frac{1}{2} \right )$.
Should flow phenomena be of interest as well,
we should just add a quadratic term in the flow speed 
to the local equilibrium distribution functions (\ref{fequil}).
In the following, we shall refer to
a three-dimensional cubic lattice of size $L \times L \times L$ 
with ${\bf c}_0=(0,0,0)$ and
${\bf c}_i=(\pm \Delta x / \Delta t,0,0)$,$(0,\pm  \Delta x / \Delta t,0)$,
$(0,0,\pm  \Delta x / \Delta t)$, $i=1,2,...,6$.
In this case $w_0=1/3$ and $w_i=1/9$ for $c_i= \Delta x / \Delta t$,
$i=1,2,...,6$. 
We enforce sterical constraints which have proven to be effective to capture
the physics of glassy systems where density is the dominant observable 
\cite{garr,rit}.
Kinetic constraints on the evolution of the system 
are enforced by the following functional rule: Propagation 
from a site ${\bf r}$ to one of its $6$ neighbors ${\bf r'}$
is permitted only if the {\it non-local} densities
$\rho_{nl}({\bf r}) =\sum_{i=1}^{6} \rho({\bf r} + {\bf c}_i \Delta t)$ and
$\rho_{nl}({\bf r'}) =\sum_{i=1}^{6} \rho({\bf r'} + {\bf c}_i \Delta t)$
prior and after streaming, respectively, both lie below a given density
threshold, $S$.
In the limit $S \rightarrow \infty$, the effective propagator
taking the system from site ${\bf r}$ at time $t$ to site
${\bf r'}$ at time $t+\Delta t$, reduces to the standard 
free-particle form, 
$G_i({\bf r},{\bf r'}; \Delta t)=\delta({\bf r'}-{\bf r}-{\bf c}_i \Delta t)$. 
In the opposite limit, $S \rightarrow 0$, no motion is allowed and
$G_i \rightarrow \delta({\bf r'}-{\bf r})$ at 
all sites, 
corresponding to structural arrest.
It is therefore clear that the 
ratio $<\rho>/S$ ($<...>$ stands for a spatial average), 
serves as a control parameter driving the system from 
the purely diffusive to the structural arrest regime.
The above rule is implemented as follows:
{\it 1.} Initialize the system by randomly 
choosing $N$ lattice sites and set them at a local density $\rho_0$. 
The remaining $L^3 - N$ sites are set at density $0$. 
The average density in the system
is then $<\rho> = \chi \rho_0 \le \rho_M$, where
$\rho_M = \rho_0$ is the maximum possible average 
density in the system, and $\chi=N/L^3$ is the concentration 
of 'loaded' sites;
{\it 2.} Compute local densities $\rho({\bf r})$ via eq.~(\ref{dens});
{\it 3.} Compute the equilibrium distribution functions $f_i^e$ via
eq.~(\ref{fequil});
{\it 4.} Perform the collision (\ref{lbe1}) on all the lattice sites
to compute $f_i^c$;
{\it 5.} Look for {\it all} the lattice sites ${\bf r}^*$ 
such that 
$\sum_{i=1}^6 \rho({\bf r}^* + {\bf c}_i) < S$
where $S$ is a fixed threshold;
{\it 6.} Perform a pre-streaming, 
according to eq.~(\ref{lbe2}), of the $f_i^c$ computed at step 4 {\it only}
along links emanating from the lattice sites ${\bf r}^*$
and pointing towards neighbor sites ${\bf r}^*$; 
{\it 7.} Compute again local densities $\rho^*({\bf r})$ 
on {\it all} the lattices sites;
{\it 8.} Look for {\it all} the lattice sites ${\bf r}^{**}$ 
such that 
$\sum_{i=1}^6 \rho^*({\bf r}^{**}+ {\bf c}_i) < S$;
{\it 9.} Perform the effective streaming step of the $f_i^c$ computed at step 4
only along links 
from ${\bf r}^*$ to ${\bf r}^{**}$ sites (these links will be denoted as
{\it active} ones), otherwise the $f_i^c$ are not moved;
{\it 10.}  Go to {\it 2.}.
The LB scheme described above is intrinsically
distinct from those used for non-ideal fluids.
Indeed, while the latter include potential energy via
effective interactions which leave the free-propagator
(kinetic energy) unaffected, in our case the interactions
alter the structure of the kinetic energy operator.

We have simulated two lattice sizes, $L=16$ and $32$, with
$\Delta x = \Delta t =1$. 
We changed $\omega$ in the range $[0.1:1]$ and did
not find any dependence of results on its specific value.
Thus, we set $\omega=0.1$.
We used $\rho_0=0.5$ and $S=1.5$, corresponding to
single site density threshold $\rho_S=S/6=0.25$. 
Keeping $\rho_0$ fixed, we chose the smallest value of $S$ 
such as to ensure sluggish dynamics
at high densities, while still allowing the system 
to be uniform at low densities.
By running several simulations with $\rho_0=0.5$, we have found that for
$S \gtrsim 1$ the system evolves by diffusive smoothing
of the density gradients
towards a long-time state characterized by a uniform density pattern 
when $<\rho> \lesssim 0.03$. 
Moreover, when $S \gtrsim 2.3$, the 
system does not show any singular behavior for densities smaller than the 
maximum possible one, $\rho_M$ (see below). This latter feature
is also observed by increasing $\rho_0$ while keeping $S=1.5$. 
This is because, by increasing $\rho_0$ at a fixed average
density $<\rho>$, the number $N$ of lattice sites to be initialized
with density $\rho_0$ decreases, and consequently it 
becomes more difficult for the kinetic constraints to be
effective. 
In conclusion, the system seems to exhibit a non-smooth transition 
from diffusive
to sluggish behavior as the reduced density 
$\lambda=<\rho>/\rho_S$ is increased.

This is shown in Fig.~1, where we plot the order parameter 
$m=(\rho_{max}-\rho_{min})/\rho_0$ as a function of $\lambda$
by keeping $<\rho>=0.12$ and $\rho_0=0.5$ fixed and varying $S$. Here,
$\rho_{max}$ and $\rho_{min}$ are the maximum and minimum values of $\rho$,
respectively, at steady-state. The values of $m$ were averaged over
50 independent runs on systems of size $L=32$.
The simulations yield $m \simeq 0$ for  
$\lambda \lesssim 0.36$, indicating a purely diffusive behavior, then 
$m$ undergoes a sharp rise and the system enters the sluggish regime.
At values of $\lambda \gtrsim 1.44$, 
the system is nearly frozen in its initial configuration
and the order
parameter $m$ settles down to $1$ since 
$\rho_{max}=\rho_0$ and $\rho_{min}=0$. 
Figure 1 clearly indicates the existence of three distinct regimes, namely
a low-density diffusive regime at $\lambda <\lambda_D \simeq 0.36$, 
a high-density frozen regime at $\lambda > \lambda_F \simeq 1.44$, 
and a sluggish regime at intermediate densities
$\lambda_D < \lambda < \lambda_F$.
Since the initial density of occupied sites is $\rho_0$, the initial 
non-local density
$\rho_{nl}({\bf r}) =\sum_{i=1}^{6} \rho({\bf r} + {\bf c}_i \Delta t)$ 
can be at most $6 \rho_0$. 
Therefore, the condition for a purely diffusive behavior 
is $6 \rho_0 \lesssim S \Rightarrow \lambda \lesssim 0.24$, 
which is in a reasonable  agreement with the value provided by simulations.
The value of $\lambda$ marking the transition
from the glassy to the frozen regime can
be determined by considering that the average initial 
non-local density is
$\rho_{nl}({\bf r}) = 6 \chi \rho_0$. When $6 \chi \rho_0 \gtrsim S \Rightarrow
\lambda \gtrsim 1.0$, the system is likely to be frozen.
This underestimates the value obtained from 
simulations, but the value of the order parameter 
$m(1) \simeq 1.3$, well below its maximum $m \simeq 2.4$, 
indicates that the system is approaching the frozen regime.
We choose $\lambda=0.48$, in order
to select a regime with clear departure from ideal fluid behavior 
($m(0.48) \simeq 2.35$).
\begin{figure}[ht]
\begin{center}
\centerline{\epsfig{file=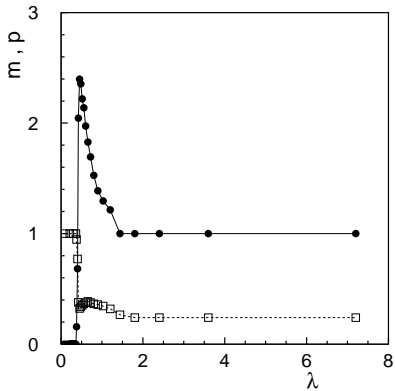,bbllx=47pt,bblly=175pt,
bburx=530pt,bbury=656pt,width=0.29\textwidth,clip=}}
\caption{The order parameter $m$ ($\bullet$) as a function of the 
reduced density $\lambda$ 
with $\rho_0 = 0.5$, $<\rho> = 0.12$, and $L=32$.
Also shown is the steady-state value of 
participation number $p$ ($\Box$), defined as 
$p = 1 / (L^3 \sum_{i=1}^{L^3} p_i^2)$, with 
$p_i = \rho({\bf r}_i)/(\chi L^3 \rho_0)$ being
the occupation probability of the site $i$. 
By definition $p(t=0)=\chi$ and $p=1$ in the case of uniform
distribution.
This plot shows that $\chi < p < 1$ 
indicating that the system
never becomes more localized than in the initial configuration.}
\end{center}
\label{param} 
\end{figure}
For this set of parameters the system 
evolves from the initial 
random configuration forming some clusters until, at long times, 
it gets arrested in one of the highly heterogeneous states. 
The arrest time decreases with increasing average density $<\rho>$.
To analyze the qualitative difference with respect to
the simple diffusion dynamics, we also
inspected the probability distribution functions of density. 
It is observed that the system, which starts from a 
two-peak distribution at $\rho=0$ and $\rho_0$, 
rapidly fills-up all available values in the range $[0:1.1]$. 
At long times there is a sharp peak at 
$\rho \simeq 0.05$, which corresponds to the background density 
and the distribution is non uniform everywhere else,
with a pronounced shoulder at $\rho \simeq 0.1$. 

We computed the time autocorrelation function 
\begin{equation}
h(t) = \frac{<<\delta \rho(t+t_0) \delta \rho(t_0)>>}{<<\delta \rho(t_0) 
\delta \rho(t_0)>>}
\end{equation}
where $<<...>>$ denotes an average over space and initial times $t_0$
and $\delta \rho(t) = \rho(t) - <\rho>$.
The plots for several values of the initial density $<\rho>$ are shown
in Fig.~2 for the case $L=32$. 
Data were obtained by averaging over 50 independent runs for each value
of $<\rho>$. For very small values of $<\rho>$, the system goes 
to a final state with uniform density and $h(t)$ relaxes to zero. 
By increasing $<\rho>$, $h(t)$ starts 
forming a plateau and stays close to the unit value for a time span 
which increases rapidly with increasing mean density $<\rho>$. 
Even at high densities, the correlator
does not show the ``two-step'' relaxation behavior 
often found in glassy materials. 
We believe that this is due the absence of a rattling motion
in our model (a similar behavior is found in the Kob-Andersen (KA) 
model \cite{KA}
for lattice glasses).
\begin{figure}[ht]
\begin{center}
\centerline{\epsfig{file=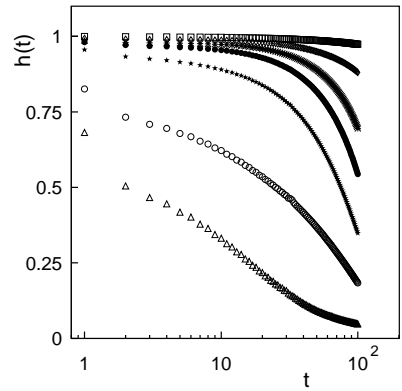,bbllx=42pt,bblly=170pt,
bburx=524pt,bbury=648pt,width=0.29\textwidth,clip=}}
\caption{Autocorrelation function $h(t)$ as a function of time $t$
for $<\rho> = 0.08 (\triangle)$,
$0.09 (\circ)$, $0.10 (\star)$, $0.11 (\bullet)$, $0.12 (\ast)$,
 $0.18 (\Diamond)$, $0.30 (\Box)$ with $\rho_0=0.5$, $S=1.5$, and $L=32$.}
\end{center}
\label{fig:correl} 
\end{figure}

We compared our results with the predictions of mode-coupling 
theory (MCT) \cite{mct}. MCT predicts a power-law time decay
away from the plateau. We therefore tried to fit the short-time behavior
of the function $h(t)$ with a power law of the form $f - B t^b$, where
$f$, $B$ and $b$ are fitting parameters.
Such a power-law decay is indeed reproduced by
our simulations in the high-density regime where from all the fits we
find $f$ very close to 1 (to within $10^{-3}$) as in the KA
model \cite{KA}. The coefficient $B$
decreases at increasing values of $<\rho>$ 
varying in the range $[4 \times 10^{-5}:3 \times 10^{-4}]$.
At variance with MCT predictions, the power-law exponent $b$ is not
independent of density but decreases at increasing values of $<\rho>$ 
varying in the range $[0.8:1.0]$. 
This density dependence of $b$ is found also in the KA model \cite{KA}
with $b$ decreasing at increasing values of $<\rho>$ 
varying in the range $[0.8:1.1]$, which is consistent with our results.
In MCT the decay at longer times is
predicted to be a stretched exponential, 
$h(t) \propto \exp{(-(t/\tau)^\beta)}$,
where the exponent $\beta$ is density-independent and
the relaxation time $\tau$ is the relevant physical parameter. Indeed, this
time scale increases strongly as density is increased (MCT predicts  
a singular behavior for $\tau$ at a density smaller than the maximum one
$\rho_M$).
Chemically different materials relax in a qualitatively 
similar manner with
relaxation functions obeying the Kohlraush-William-Watts function
$\exp{(-(t/\tau)^\beta)}$ \cite{KWW}.
We fitted successfully $h(t)$ at long times for $<\rho> \ge 0.12$
by using a stretched exponential with $\tau$ and $\beta$ as fitting 
parameters
and found that the inverse relaxation time $1/\tau$ 
vanishes at a critical density $\rho_c = 0.412 \pm 0.010$, 
with power law $6(\rho_c - <\rho>)^{\gamma_c}$,
being $\gamma_c=4.68$ (in the KA model the critical exponent 
$\gamma_c$ is $\sim 5$ \cite{KA}),
while the stretching exponent $\beta$ is density-dependent. 
MCT predicts such a power law behavior for the relaxation time
with a system-dependent exponent $\gamma_c$ \cite{GS}.
It is remarkable that, despite the density dependence, we find 
numerically that $\beta(<\rho>) \simeq b(<\rho>)$, as predicted in MCT
for the time dependence of the autocorrelation function \cite{fuchs}.
We stress that the critical value $\rho_c$ is smaller than the maximum
density $\rho_M$, corresponding to a fully-loaded lattice. 
The plot of the relevant physical quantity $1/\tau$ as a function of 
$\rho_c - <\rho>$ is shown in Fig.~3, where we also report 
the fitting values of $1/\tau$ for the system size $L=16$. 
Also in this case, we found that $1/\tau$ vanishes with a power-law behavior
and the estimated critical density is $0.405 \pm 0.029$, which is consistent
with $\rho_c$ within the error range, with no significant lattice
size effect. This suggests a singular behavior at about $<\rho> = 0.412$. 
\begin{figure}[ht]
\begin{center}
\centerline{\epsfig{file=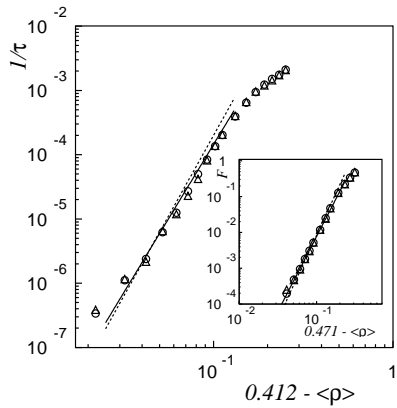,bbllx=29pt,bblly=156pt,
bburx=527pt,bbury=675pt,width=0.29\textwidth,clip=}}
\caption{Inverse relaxation time $1/\tau$ as a function of $0.412 -$ $<\rho>$
for $\rho_0=0.5$, $S=1.5$ and $L= 16 (\triangle)$, $32 (\circ)$.
The full line has slope $4.68$.
In the inset the fraction $F$ of active links versus $0.471 - <\rho>$ 
is shown for the same parameters. 
The full line has slope $4.19$. For a comparison we plotted 
the results of the KA model by using dashed lines.}
\end{center}
\label{fig:tau} 
\end{figure}
We also inspected the ratio $F$ of active lattice links 
to the 
total number $6 L^3$ of lattice links. 
This quantity can be viewed as a measure of the 
degree of glassiness, as observed in kinetically constrained 
lattice glass models \cite{bertin}, since we expect the
relaxation time $\tau$ to diverge in the limit $F \rightarrow 0$.
In the limit $<\rho> \rightarrow 0$, $F \rightarrow 1$ since
the kinetic constraints do not play any role. 
When  $<\rho> \rightarrow \rho_c$, we expect $F \rightarrow 0$ 
since in this limit the kinetic constraints 
are very effective in slowing down the dynamics \cite{KA}. 

In the inset of Fig.~3 we plot the steady-state values 
of $F$, averaged over 50 runs, for each value of $<\rho>$. 
We found that $F$ vanishes at a critical density 
$\rho_c^{'} = 0.471 \pm 0.005$, with power law 
$120(\rho_c^{'} - <\rho>)^{\gamma_c^{'}}$, being $\gamma_c^{'}=4.19$. 
In the case of the KA model, a similar power law is found
with critical exponent $\gamma_c^{'} \sim 4.7$ \cite{KA}. 
In the same inset we also report 
the values of $F$ for the system size $L=16$. 
Also in this case we found that $F$ vanishes with a power law behavior
and the estimated critical density is $0.482 \pm 0.016$, which is consistent
with $\rho_c^{'}$ within the error range, and no significant lattice
size effect.
It is interesting to note that 
$\rho_c^{'} \simeq \rho_c$, supporting
the conclusion that our model shows singular behavior 
at a density smaller than the maximum possible one, $\rho_M$. 

Summarizing, we have introduced 
a mesoscopic LB model 
which appears to reproduce {\it some} physical features 
of dynamically heterogeneous fluids, such as
sluggish relaxation and continuum density distributions.
To this purpose, the standard LB
dynamics has been augmented with self-consistent constraints 
based on the non-local density of the surrounding fluid.
A typical run on a $32^3$ lattice (four times larger than
typical lattice glass simulations \cite{KA,coniglio}) takes just 
a few minutes on a 2.4 GHz Intel Xeon processor. 

\begin{acknowledgments}
Illuminating discussions with K. Binder, W. Kob, E. Marinari and
G. Parisi are kindly acknowledged.
\end{acknowledgments}

\end{document}